\documentstyle[12pt,twoside]{article}

\normalbaselines
\font\tbf = cmbx12

\pagebreak

\begin{document}

\indent
\vskip 0.1cm
\centerline{\tbf TERNARY  ALGEBRAIC  STRUCTURES}
\vskip 0.3cm
\centerline{\tbf AND  THEIR  APPLICATIONS  IN  PHYSICS}
\vskip 0.7cm
\centerline{\tbf Richard Kerner}
\vskip 0.5cm
\centerline{\it Laboratoire GCR - Universit\'e Pierre-et-Marie-Curie,}
\vskip 0.2cm
\centerline{\it Tour 22, 4-\`eme \'etage, Bo{\i}te 142, 4 Place Jussieu, 75005 Paris}
\vskip 0.2cm
\centerline{{\it e-mail :} \,  rk@ccr.jussieu.fr}
\vskip 0.7cm
\centerline{\tbf Abstract.}  
\vskip 0.3cm
\indent
\small{We discuss certain ternary algebraic structures appearing more or less
naturally  in various domains of theoretical and mathematical physics. Far
from being exhaustive, this article is intended above all to draw attention
to these algebras, which may find more interesting applications in the years
to come.}
\vskip 0.7cm
\indent
{\tbf 1. INTRODUCTION.}
\vskip 0.3cm
\indent
Ternary algebraic operations and cubic relations have been considered,
although quite sporadically, by several authors already in the XIX-th century,
e.g. by A. Cayley (\cite{Cayley}) and J.J. Sylvester ( \cite{Sylvester}. The
development of Cayley's ideas, which contained a cubic generalization of 
matrices and their determinants, can be found in a recent book by M. Kapranov,
I.M. Gelfand and A. Zelevinskii (\cite{Kapranov}). A discussion of the next
step in generality, the so called $n-ary$ algebras, can be found in
(\cite{Vainerman}). Here we shall focus our attention on the {\it ternary}
and {\it cubic} algebraic structures only.
\newline
\indent
We shall introduce the following distinction between these two denominations:
we shall call a {\it ternary algebraic structure} any linear space $V$
endowed with one or more ternary composition laws:
$$m_3 : \, \, V \otimes V \otimes V \Rightarrow V \, \ \ \, \ \ \, {\rm or}
\, \ \ \, \, m'_3 : \, \ \ V \otimes V \otimes V \Rightarrow {\bf C} \, , $$
the second law being an analogue of a scalar product in the usual (binary)
case.
\newline
\indent
We shall call a {\it cubic structure} or an algebra generated by cubic
relations, an ordinary algebra with a binary composition law:
$$ m_2 \, : \, \ \ \, V \otimes V \Rightarrow \,  V $$
with cubic (third order) defining relations for the generators: e.g.
$(abc)= e^{2 \pi i/3} \, (bca) $
\newline
\indent
Some of ternary operations and cubic relations are so familiar that we don't
pay much attention to their special character. We can cite as example the
triple product of vectors in 3-dimensional Euclidean vector space:
$$ \{a,b,c \} = \vec{a} \cdot (\vec{b} \times \vec{c} ) =
\vec{b} \cdot (\vec{c} \times \vec{a} ) $$
which is a tri-linear mapping from $E \otimes E \otimes E \, \ \ $ onto
$\bf{R}^1$., invariant under the cyclic group $Z_3 \, .$
\newline
\indent
Curiously enough, it is in the 4-dimensional Minkowskian space-time $M_4$
where a natural ternary composition of 4-vectors can be easily defined:
$$ (X,Y,Z) \rightarrow U(X,Y,Z) \in M_4 $$
with the resulting 4-vector $U^{\mu}$ defined via its components in a given
coordinate system as follows:
$$U^{\mu} \, (X,Y,Z) = g^{\mu \sigma} \, \eta_{\sigma \nu \lambda \rho} \,
X^{\nu} Y^{\lambda} Z^{\rho} , \, \ \ \, \ \ {\rm with } \, \ \ \,
\mu, \nu , ... = 0,1,2,3 .$$
where $g^{\mu \nu} $ is the metric tensor, and $\eta_{\mu \nu \lambda \rho}$
is the canonical volume element of $M_4 . $
\newline
\indent
Other examples of ``ternary ideas'' that we should cite here are:
\vskip 0.2cm
\indent
\hskip 0.5cm
- {\it cubic matrices} and a generalization of the determinant, called the
{\it ``hyperdeterminant''}, first introduced by Cayley in 1840, then found
again and generalized by Kapranov, Gelfand and Zelevinskii in 1990
(\cite{Kapranov}). The simplest example of this (non-commutative and
non-associative) ternary algebra is given by the following composition rule:
$$ \{ a \, , b \, , c \}_{ijk} = {\displaystyle{\sum_{l, m, n} } \, a_{n i l}
\, b_{l j m} \, c_{m k n} } , \, \ \ \, \ \ i, j , k ... = 1, 2, ..., N$$
Other ternary rules can be obtained from this one by taking various linear
combinations, with real or complex coefficients, of the above 3-product, e.g.
$$ [a, b, c] = \{ a,b,c \} + \omega \, \{ b,c,a \} + \omega^2 \, \{ c,a,b \} \, \
\ \,  \ \  \,  {\rm with} \ \ \, \omega = e^{ 2 \pi i/3} \, . $$
\vskip 0.2cm
\indent
\hskip 0.5cm
- the algebra of ``{\it nonions}'' , introduced by Sylvester as a ternary
analog of Hamilton's quaternions. The ``nonions'' are generated by two matrices:
$$ \eta_1 = \pmatrix{0&1&0 \cr 0&0&1 \cr 1&0&0} \, \ \ \, \eta_2 =
\pmatrix{0 & 1 & 0 \cr 0 & 0 & \omega \cr \omega^2 & 0 & 0 } $$
and all their linearly independent powers; the constitutive relations are of
cubic character:
$$ \displaystyle{\sum_{perm. (ikm)}} \Gamma_i \Gamma_k \Gamma_m = \delta_{ikm}
\, {\bf 1} $$
where $\delta_{ikm}$ is equal to $1$ when $i=k=m$ and $0$ otherwise.
\vskip 0.2cm
\indent
\hskip 0.5cm
- cubic analog of Laplace and d'Alembert equations, first considered  by
Himbert (\cite{Himbert}) in 1934: the third-order differential operator that
generalized the Laplacian was
$$ (\frac{\partial}{\partial x} +  \frac{\partial}{\partial y} +
\frac{\partial}{\partial z} )
\,  (\frac{\partial}{\partial x} + \omega \frac{\partial}{\partial y} + \omega^2
\frac{\partial}{\partial z} )
 \, (\frac{\partial}{\partial x} + \omega^2 \frac{\partial}{\partial y} + \omega
\frac{\partial}{\partial z} ) $$
$$= \frac{\partial^3}{\partial x^3} + \frac{\partial^3}{\partial y^3} +
\frac{\partial^3}{\partial z^3} - 3 \, \frac{\partial^3}{\partial x \partial
y \partial z } $$
\indent
Other ternary and cubic algebras have been studied by Ruth Lawrence, L. Dabrowski,
F. Nesti and P. Siniscalco (\cite{Dabrowski}), Plyushchay and
Rausch de Traubenberg ({\cite{Plyushchay}), and other authors.

\newpage
\indent
{\tbf 2. IMPORTANT  TERNARY  RELATIONS  IN  PHYSICS.}
\vskip 0.3cm
\indent
The quark model inspired a particular brand of ternary algebraic systems,
intended to explain the non-observability of isolated quarks as a phenomenon
of ``algebraic confinement''.  One of the first such attempts has been
proposed by Y. Nambu (\cite{Nambu1}) in 1973, and known under the name of
``{\it Nambu mechanics}'' since then (\cite{Takhtajan}).
\newline
\indent
Consider a 3-dimensional real space parametrized by Cartesian coordinates,
with $ \vec{r} = (x,y,z) \in {\bf R}^3$. Introducing two smooth functions
$H(x,y,z)$ and $G(x,y,z)$, one may define the following ternary analog of the
Poisson bracket and dynamical equations : for a given function $f(x,y,z)$
defined on our 3-dimensional space, its time derivative is postulated to be:
\begin{equation}
\frac{d {\vec{r}}}{dt} = (\overrightarrow{\nabla} H) \times
(\overrightarrow{\nabla} G )
\end{equation}
or more explicitly, because we have
$$ \frac{dx}{dt} = {\rm det} \, \biggl( \frac{\partial (H,G)}{\partial (y,z)}
\biggr) \, , \ \  \,
 \frac{dy}{dt} = {\rm det} \, \biggl( \frac{\partial (H,G)}{\partial (z,x)}
 \biggr) \, , \ \  \,
 \frac{dz}{dt} = {\rm det} \, \biggl( \frac{\partial (H,G)}{\partial (x,y)}
 \biggr) \, ,  $$
we can write
\begin{equation}
\frac{d f}{dt} = (\overrightarrow{\nabla}) \cdot (\overrightarrow{\nabla} H )
\times (\overrightarrow{\nabla} G) =  {\rm det} \, \biggl(
\frac{\partial(f,G,H)}{\partial(x,y,z)} \biggr) =    [f, H, G ]
\end{equation}
The so defined {\it ``ternary Poisson bracket''} satisfies obvious relations:
$$ a) \, \ \ [A,B,C] = - [B,A,C] = [B,C,A] \, ; $$
$$ b) \, \ \ [A_1 \, A_2 , B, C] = [A_1,B,C] \, A_2 + A_1 \, [A_2,B,C] \, ; $$
$$ c) \, \ \ \overrightarrow{\nabla} \cdot \biggl( \frac{d \vec{r}}{dt} \biggr)
= \overrightarrow{\nabla} \cdot ( \overrightarrow{\nabla} H \times
\overrightarrow{\nabla} G ) = 0 \, . $$
A canonical transformation $(x,y,z) \Rightarrow (x',y',z')$ is readily defined
as a smooth coordinate transformation whose determinant is equal to $1$:
$$ [x',y',z']= {\rm det} \, \biggl(\frac{\partial(x',y',z')}{\partial(x,y,z)}
\biggr) = 1
\, ,$$
$$ {\rm so \, \,  that \, \,  one \, \ automatically \, \, has \, : \ \ \, }
\, \ \ \, \ \ \, \frac{d f}{dt} = {\rm det} \,
\biggl( \frac{\partial(f,H,G)}{\partial(x,y,z)} \biggr) = {\rm det} \,
\biggl( \frac{\partial(f,H,G)}{\partial(x',y',z')} \biggr) \, \ \ \, \ \ \, $$
\indent
It is easily seen that {\it linear canonical transformations} leaving this
ternary Poisson bracket invariant form the group $SL(3, {\bf R})$.
\newline
\indent
The dynamical equations describing the {\it Euler top} can be cast into this
new ternary mechanics scheme, if we identify the vector $\vec{r}$ with the
components of the angular momentum $\vec{L} = [ L_x, L_y, L_z ] $, and the
two ``Hamiltonians'' with the following functions of the above:
\begin{equation}
H = \frac{1}{2} \, \biggl[ L^2_x + L^2_y + L^2_z \biggr] , \, \ \ \, \ \
G = \frac{1}{2} \, \Biggl[ \frac{L^2_x}{J_x} + \frac{L^2_y}{J_y} +
\frac{L^2_z}{J_z} \Biggr].
\end{equation}
Recently R. Yamaleev has found an interesting link between the Nambu mechanics
and ternary $Z_3$-graded algebras (\cite{Yamaleev1}).
\newpage
\indent
The Yang-Baxter equation provides another celebrated
cubic relation imposed
on the bilinear operators named $\tilde{R}$-matrices : for
${\tilde{ R}}_{km} : V \otimes V \rightarrow V \otimes V ,$ one has
\begin{equation}
{\tilde{R}}_{23} \circ {\tilde{R}}_{12} \circ {\tilde{R}}_{23} =
{\tilde{R}}_{12} \circ {\tilde{R}}_{23} \circ {\tilde{R}}_{12} ,
\end{equation}
where the indeces refer to various choices of {\it two} out of {\it three}
distinct specimens of the vector space $V$.
\newline
\indent
An alternative formulation of this formula is more widely used. Let $P$ be
the operator of permutation, $P : V_1 \otimes V_2 \rightarrow V_2 \otimes V_1 $
and let us introduce another $R$-matrix by defining ${\tilde{R}} = P \circ R$.
Then the same relation takes on the following form:
\begin{equation}
R_{12} \circ R_{13} \circ R_{23} =  R_{23} \circ R_{13} \circ R_{12} .
\end{equation}
Applications of this equation are innumerable indeed; they serve to solve
many integrable systems, such as Toda lattices; they also give the
representations of braid groups, etc.
\newline
\indent
In a given local basis of $V \otimes V , \, \ \ e_i \otimes e_k$, we can
write, for $X= X^i \, e_i, \, Y = Y^k \, e_k$,
$$ R(X,Y) = R^{km}_{\, \ \ \, ij} \, X^i Y^j \, e_k \otimes e_m $$
\indent
An interesting ternary aspect of these $R$-matrices has been discovered by
S. Okubo (\cite{Okubo1}) in search for new solutions of Yang-Baxter equations.
Introducing a supplementary real parameter  $\theta$ , we can write this
equation as follows:
\begin{equation}
R^{b' \, a'}_{\, \ \ \, \, a_1 \, b_1} \, (\theta) \, R^{c' \, a_2}_{\, \ \ \,
\, a' \, c_1} \, (\theta') \, R^{c_2 \, b_2}_{\, \, \ \ \, b' \, c'} \,
(\theta'') =
R^{c' \, b'}_{\, \ \ \, \, b_1 \, c_1} \, (\theta'') \, R^{c_2 \, a'}_{\, \ \ \,
\, a_1 \, c'} \, (\theta') \, R^{b_2 \, a_2}_{\, \, \ \ \, a' \, b'} \, 
(\theta)  \, ,
\end{equation}
with $\theta' = \theta + \theta''$.
\newline
\indent
An entire class of solutions of Yang-Baxter equation, including the ones found
by de Vega and Nicolai, can be obtained in terms of {\it triple product systems}
if the matrix $R$ satisfies an extra
symmetry condition:
\begin{equation}
R^{b \, a}_{ \, \ \ \, \, d \, c} \, (\theta) =
R^{a \, b}_{\, \ \ \, \, c \, d} \, (\theta).
\end{equation}
\indent
Okubo considered the following {\it symplectic} and {\it orthogonal} triple
systems, i.e. vector spaces (denoted by $V$) endowed simultaneously with a
non-degenerate bi-linear form
$$ \langle x, y \rangle \, : \, \ \ V \otimes V \rightarrow {\bf C}^1 \, , \, \ \ \,
x, y \in V $$
$$ {\rm and \, \, a \, \, triple \, \, product  \, \ \ \, \ \ \ \ \, \ \ \, }
\{ x, y , z \} \, : \, \ \ V \otimes V \otimes V \rightarrow V \, , \, \ \
\, x,y,z \in V  \, \ \ \, \ \ \, \ \ \, \ \ $$
The fundamental assumptions about the relationship between these two products
are :
$$ a) \, \langle y,x \rangle = \varepsilon \langle x,y \rangle ; \, \ \ \,
b) \, \{ y, x ,z \} = - \varepsilon \, \{ x, y, z \} \, , $$
\indent
If $\varepsilon = -1$, the system is called {\it symplectic}; if $\varepsilon
= 1$, it is called {\it orthogonal.}
$$ \, c) \, \, \ \ \,  \langle \{ u,v,x \} , y \rangle = - \langle x,
\{ u, v, y \} \rangle \, ;  $$
$$ \, d) \, \, \ \ \, \{ u, v \{ x,y,z\} \} = \{ \{ u,v,x \}, y,z \} +
\{ x , \{ u,v,y \} , z \} + \{ x, y , \{ u,v,z \} \} $$
$$ \, e) \, \ \ \, \, \{ x,y,z \} + \varepsilon \, \{ x,z,y \} = 2 \,
\lambda_0 \, \langle y,z \rangle \, x - \lambda_0 \, \langle x,y \rangle \,
z - \lambda_0 \, <z,x> \, y . $$
with a free real parameter $\lambda_0$.
\newpage
\indent
In a chosen basis of $V , \, \ \ ( e_1 , e_2 , ... , e_N )$, one can write
$$ \langle e_i , e_k \rangle = g_{ik} = \varepsilon \, g_{ki}, \, \ \ \,
{\rm and} \, \ \ \, \ \ \{e_i , e_k, e_m \} = C^{\,j}_{\, \ \ ikm} \, e_j $$
where the coefficients $C^{\,j}_{\, \ \ ikm}$ play the r\^ole of
ternary structure constants.
\newline
\indent
With the help of the inverse metric tensor, $g^{jk}$, we can now raise the
lower-case indeces, defining the contravariant basis $e^k = g^{km} \, e_m$.
If a one-parameter family of triple products is defined,
$\{e_i,e_k,e_m \}_{\theta}$, then we may define an $R$-matrix depending on
the same parameter $\theta$:
$$ R^{ij}_{\, \ \ \, \ \ k m } = \langle e^i , \{e^j, e_k ,e_m \}_{\theta}
\, \rangle $$
or equivalently,
\begin{equation}
\{ \, e^b, \, e_c \, e_d \, \}_{\theta} = R^{ab}_{\, \ \  \, \, cd} \, e_a .
\end{equation}
The symmetry condition $R^{b \, a}_{ \, \ \ \, \, d \, c} \, (\theta) =
R^{a \, b}_{\, \ \ \, \, c \, d} \, (\theta) $ can be now written as
$$ \langle \, u , \{ x,y,z \}_{\theta} \, \rangle = \langle \, z ,
\{ y, x, u \}_{\theta} \, \rangle $$
and the Yang-Baxter equation becomes equivalent with an extra condition
imposed on the ternary product:
\begin{equation}
{\displaystyle{\sum_a }} \, \, \{ v , \{ u , e_a , z \}_{{\theta}'} \,
, \{ e^a , x , y \}_{\theta} \, \}_{{\theta}"} =
{\displaystyle{\sum_a }} \, \, \{ u , \{ v , e_a , x \}_{{\theta}'} \, ,
\{ e^a , z , y \}_{{\theta}"} \, \}_{\theta}
\end{equation}
\indent
Using thus encoded form of the Yang-Baxter equation S. Okubo was able to
find a series of new solutions just by finding 1-parameter families of
ternary products satisfying the above constraints.
\newline
\indent
This original approach suggests another possibility of introducing ternary
staructures in the very fabric of traditional quantum mechanics. As we know,
any bounded linear operator acting in Hilbert space ${\cal{H}}$ can be
represented as
\begin{equation}
A : \, \, {\cal{H}} \rightarrow {\cal{H}} , \, \ \ \, \ \ \, \ \
A = {\displaystyle{\sum_{k,m}}} \, a_{km} \, \mid e_k \rangle \langle e_m
\mid \, ;
\end{equation}
where $\mid e_k \rangle $ is a basis in Hilbert space. If now
$$ \mid x \rangle =  {\displaystyle{\sum_k}} \, c_m \, \mid e_m \rangle , $$
then one has
\begin{equation}
A  \mid x \rangle  = \displaystyle{\sum_{i,k,m}}  a_{ik} \mid e_i \rangle
\langle e_k \mid c_m  \mid e_m \rangle = \displaystyle{\sum_{i,k,m}} a_{ik}
c_m \delta_{km}  \mid e_i \rangle = \displaystyle{\sum_{km}} a_{km}  c_m \,
\mid e_m \rangle 
\end{equation}
Each item in this sum can be cosidered as a result of {\it ternary
multiplication} defined in the Hilbert space of states:
\begin{equation}
m ( \mid e_i \rangle , \mid e_j \rangle , \mid e_k \rangle ) = \mid e_i \rangle
\langle e_j \mid e_k \rangle = \delta_{jk} \mid e_i \rangle
= \displaystyle{\sum_{n}} \delta_{jk} \delta^n_i \mid e_n \rangle ,
\end{equation}
with the structure constants defined as $C^n_{\, \ \ ijk} = \delta_{jk}
\delta^n_i \, . $
Using this interpretational scheme, the states and the observables (operators)
are no more separate entities, but can interact with each other: by superposing
triplets of states, we arrive at the result which amounts to changing both
the state and the observable simultaneously.
\newpage
\indent
Similar constructions, often referred to as {\it algebraic confinement}, were
considered by many authors, in particular by H.J. Lipkin quite a long time
ago (\cite{Lipkin}).
\newline
\indent
Consider an algebra of operators ${\cal{O}}$ acting on a Hilbert space
${\cal{H}}$ which is a free module with respect to the algebra ${\cal{O}}$,
endowed with Hilbertian scalar product. Let us introduce tensor products of
the algebra and the module with the following $Z_3$-graded matrix algebra
${\cal{A}}$ over the complex field ${\bf C}^1 $ :
$$ {\cal{A}} = {\cal{A}}_0 \oplus {\cal{A}}_1  \oplus {\cal{A}}_2 \, , \, \ \ 
\, \ \ \, {\cal{A}} \in {\rm Mat} \, (3, {\bf C} ) \, . $$
The three linear subspaces of ${\cal{A}}$, of which only ${\cal{A}}_0$ forms
a subalgebra, are defined as follows:
$$ {\cal{A}}_0 := \pmatrix{ \alpha & 0 & 0 \cr 0 & \beta & 0 \cr 0 & 0 &
\gamma} \, , \, \ \
{\cal{A}}_1 := \pmatrix{0 & \alpha & 0 \cr 0 & 0 & \beta \cr \gamma & 0 & 0 }
\, , \, \ \ {\cal{A}}_2 := \pmatrix{0&0& \gamma \cr \beta & 0 & 0 \cr 0 &
\alpha & 0 } \, . $$
\indent
It is easy to check that under matrix multiplication the degrees $0,1$ and $2$
add up modulo $3$: a product of two elements of degree $1$ belongs to
${\cal{A}}_2$, the product of two elements of degree $2$
belongs to ${\cal{A}}_1$, and the product of an element of degree $1$ with an
element of degree $2$ belongs to ${\cal{A}}_0$, etc.
\newline
\indent
With this in mind, a generalized state vector can now belong to one of these
subspaces, e.g., $ \mid \Psi \rangle$ of degree one, its hermitian conjugate
being automatically of $Z_3$-degree $2$:
$$\mid \Psi \rangle := \pmatrix{0& \mid \psi_1 \rangle & 0 \cr 0 & 0 & \mid
\psi_2 \rangle \cr \mid \psi_3 \rangle & 0 & 0 } \, ; \, \, \ \ \langle \Psi
\mid := \pmatrix{ 0 & 0 & \langle \psi_3 \mid \cr \langle \psi_1 \mid & 0 & 0
\cr 0 & \ \langle \psi_2 \mid & 0 } $$
with $ \mid \psi_k \rangle \in {\cal{H}}$.
The scalar product obviously generalizes as follows:
$$ \langle \Phi \mid \Psi \rangle := Tr  \Biggl[ \pmatrix{ 0 & 0 & \langle
\phi_3 \mid \cr \langle \phi_1 \mid & 0 & 0 \cr 0 & \ \langle \phi_2 \mid
& 0 }  \, \pmatrix{0& \mid \psi_1 \rangle & 0 \cr 0 & 0 & \mid \psi_2 \rangle
\cr \mid \psi_3 \rangle & 0 & 0 } \Biggr] = $$
\begin{equation}
= \langle \phi_1 \mid \psi_1 \rangle + \langle \phi_2 \mid \psi_2 \rangle +
\langle \phi_3 \mid \psi_3 \rangle \, .
\end{equation}
\indent
With this definition of scalar product any expectation value of an operator
of degree $1$ or $2$ ( represented by the corresponding traceless matrices )
will identically vanish, e.g. for an operator of degree $1$ : $ ({\cal{D}}_i
\in {\cal{O}} )$
$$ Tr \Biggl[ \pmatrix{ 0 & 0 & \langle \psi_3 \mid \cr
\langle \psi_1 \mid & 0 & 0 \cr 0 & \ \langle \psi_2 \mid & 0 }  \,
\pmatrix{ 0 & D_1 & 0 \cr 0 & 0 & D_2 \cr D_3 & 0 & 0 } \,
\pmatrix{ 0 & \mid \psi_1 \rangle & 0 \cr 0 & 0 & \mid \psi_2 \rangle \cr
\mid \psi_3 \rangle & 0 & 0 } \Biggr] = 0 . $$
It is clear that only the operators whose $Z_3$-degree is $0$ may have
non-vanishing expectation values, because the operators of degrees $1$ and
$2$ are traceless. Denoting  the operator of degree $1$ by $Q$, and the
operators of degree  $2$ by $\bar{Q}$, the only combinations that can be
observed, i.e. that can lead to non-vanishing expectation values no matter
what the nature of the operator and the observable it is supposed to represent,
are the following products:
\begin{equation}
Q \, Q \, Q \, ; \ \ \, \ \ \,  \bar{Q} \, \bar{Q} \, \bar{Q} \, ;
\, \ \  \ \, Q \, \bar{Q} \, \ \ \, {\rm and} \, \ \ \, \ \, \bar{Q} \, Q 
\end{equation}
which correspond to the observable combinations (tensor products) of the
fields supposed to describe the quarks. This particular realisation of
{\it ``algebraic confinement''} suggests the importance of ternary and cubic
relations in algebras of observables.
\newpage
\indent
{\tbf 3. CUBIC GRASSMANN AND CLIFFORD ALGEBRAS.}
\vskip 0.3cm
\indent
A general $3$-algebra (or {\it ternary algebra}) is defined as internal ternary
multiplication in  a vector space $V$. Such a multiplication must be of course
$3$-linear, but not necessarily associative:
\begin{equation}
m : \, \ \  V \otimes V \otimes V  \rightarrow V \, ; \, \ \ \,
m(X,Y,Z) \in V
\label{3product}
\end{equation}
Such a $3$-product is said to be  {\it strongly associative} if one has
\begin{equation}
m(X,m(S,Y,T),Z) = m(m(X,S,Y),T,Z) = m(X,S, m(Y,T,Z))
\end{equation}
\indent
Of course, any associative binary algebra can serve as starting point for
introduction of a (not necessarily associative) ternary algebra, by defining
its ternary product :
$$ \, \ \ \, \ \  (*) \, \ \ \, \ \ (X,Y,Z) = XYZ  \ \ \, \, \ \ \ \ \  \ \,
\ \   \, \ \ \, \ \ \, \ \ \, \ \  \, \ \ \, \ \ \, \ \  \, \ \ \, \ \ \,
\ \  
({\rm trivial }); $$
$$ \, \ \ \, \ \  (**) \, \ \ \, \ \ \{ X,Y,Z \} = XYZ + YZX + ZXY \, \ \ \,
\ \ \, \ \ \, \ \ \,  \ \ \, \ \ ({\rm symmetric} ) ;  $$
\begin{equation}
\, \ \  \, \ \ (***) \, \ \ \, \ \ [ X,Y,Z ] = XYZ + \omega \, YZX +
\omega^2 \, ZXY
\ \ \, \,
({\rm {\it \omega}-skewsymmetric})
\end{equation}
where we set $\omega = e^{2 \pi i /3}$, the primitive cubic root of unity.
It is worthwhile to note that the last cubic algebra, which is a direct
generalization of the $Z_2$-graded skew-symmetric product $[X,Y] = XY - YX ,$
which defines the usual Lie algebra product, contains it as a special
substructure if the underlying associative algebra is unital. Indeed, if
${\bf 1}$ is the unit of that algebra, one easily checks that substututing it
in place of the second factor of the skew-symmetric ternary product, one gets:
\begin{equation}
\{ X, {\bf 1} , Z \} := X {\bf 1} Z + \omega \, {\bf 1} Z X + \omega^2
Z X {\bf 1} = X Z + (\omega + \omega^2) ZX = XZ - ZX \, ,
\end{equation}
because of the identity $\omega + \omega^2 + 1 = 0$, so that the usual
Lie-algebraic structure is recovered as a special case.
\newline
\indent
In general, a ternary algebra can not be derived from an associative binary
algebra. Indeed, suppose that we have, on one side, a ternary multiplication
law defined by its structural constants with respect to a given basis
$ \{ e_k \} \, :$
$$m \, (e_i , e_j , e_k )= {\displaystyle{\sum_{l=1}^N}} \, m^l_{ \ \ ijk }
\, e_l \, ,$$
and on the other hand, a binary multiplication law, defined in the same
basis  by
$$p(e_i ,e_k)={\displaystyle{\sum_{m=1}^N}} \, p^m_{\, \ \ ik} \,e_m \,;$$
and suppose that we want to interpret the ternary multimlication as two
consecutive binary multiplications:
$$ m( a, b, c ) = p(a , p(b,c)) = p(p(a,b),c) $$
(supposing that the binary algebra is associative).
Then, after projection on the basis vectors $e_k$ we should have
\begin{equation}
m^i_{ \ \ jkm} = {\displaystyle{\sum_{r=1}^N}} \, p^r_{ \ \ km} \,
p^i_{  \ \  jr} \, .
\end{equation}
Even in the simplest case of dimension $N=2$, we get $2^4 = 16$ equations
for $2^3 = 8$ unknowns (the coefficients $p^i_{ \ \ jk}$), which can
not be solved in general, except maybe for some very special cases.
\newpage
\indent
Recently, A. Sitarz (\cite{Sitarz}) proved that any {\it associative} $n$-ary
algebra can be generated by a part of the ${\cal{A}}_1$ , i.e. the grade  $1$
subspace of certain $Z_{N-1}$-graded associative ordinary (binary) algebra.
The simplest example of this situation is given by the groups algebra of
the symmetry group $S_3$. It contains two subspaces, which are naturally
$Z_2$-graded. The even subspace (of degree 0) is spanned by the cyclic subgroup
$Z_3$, while the odd subspace is spanned by three involutions, corresponding
to odd permutations. As the square of each involution is the unit element,
the product of three involutions gives another involution, which defines a
ternary algebra (without unit element ). The full ternary multiplication
table contains 27 independent products.
\newline
\indent
Just as binary products can be divided into different classes reflecting
their behavior under the permutation group $Z_2$, so all ternary products can
be divided into classes according to their behavior under the actions of
the permutation group $S_3$. These in turn are naturally separated into
{\it symmetric cubic} and {\it skew-symmetric cubic} subsets.
\newline
\indent
There are four possible ternary generalizations of the {\it symmetric}
binary product :
$$S_0 \, : x^j x^k x^m = x^{\pi(j)} x^{\pi(k)} x^{\pi(m)} \, , \, \ \ \, \ \
{\rm any \, \ \ permutation \, \ \ \, } \pi \in {\cal{S}}_3 \, ;$$
$$S_1 \, : x^j x^k x^m = x^k x^m x^j \, \ \ \, \ \ ({\rm cyclic \, \ \
permutations \, \ \ only}) \, ;$$
$$S \, : \, \ \ \, x^k x^m x^n + \omega \, x^m x^n x^k + \omega^2 \, x^n x^k x^m = 0
\, ; $$
$$\bar{S} \, : \, \ \ \, {\bar{x}}^k {\bar{x}}^m {\bar{x}}^n + \omega^2 \, 
{\bar{x}}^m  {\bar{x}}^n  {\bar{x}}^k + \omega \, 
{\bar{x}}^n {\bar{x}}^k {\bar{x}}^m = 0 \, . $$
Obviously, the spaces $S$ and $\bar{S}$ are isomorphic, and there exist surjective
homomorphisms from $S$ and $\bar{S}$ onto $S_1$, and a surjective homomorphism
from $S_1$ onto $S_0$.
Similarly, the {\it skew-symmetric} cubic algebras can be defined as a direct
generalisation of Grassmann algebras:
$$\Lambda_0 \, : \, \theta^A \, \theta^B \, \theta^C  + \theta^B \, \theta^C \,
\theta^A  + \theta^C \, \theta^A \, \theta^B + \theta^C \, \theta^B \, \theta^A
+ \theta^B \, \theta^A \, \theta^C + \theta^A \, \theta^C \, \theta^B = 0 \,  ,$$
$$\Lambda_1 \, :\ \ \, \ \   \theta^A \, \theta^B \, \theta^C + \theta^B \, \theta^C \ \, \theta^A + \theta^C \, \theta^A \, \theta^B = 0 \, , $$
$$\Lambda \, : \, \ \ \theta^A \, \theta^B \, \theta^C = \omega \, \theta^B \, 
\theta^C \, \theta^A \, ; \, \ \ \, \ \ 
{\bar{\Lambda}} \, : \, \ \ {\bar{\theta}}^{\bar{A}} \, {\bar{\theta}}^{\bar{B}} \, {\bar{\theta}}^{\bar{C}} = \omega^2 \, 
{\bar{\theta}}^{\bar{B}} \, {\bar{\theta}}^{\bar{C}} \, {\bar{\theta}}^{\bar{A}} \, .$$
\indent
Here again, a surjective homomorphism exists from $\Lambda_0$ onto $\Lambda_1$,
then two surjective homomorphisms can be defined from  $\Lambda_1$ onto $\Lambda$
or onto $\bar{\Lambda}$
\newline
\indent
The natural $Z_3$-grading attributes degree $1$ to variables $\theta^A$ and
degree $2$ to the variables ${\bar{\theta}}^{\bar{B}}$ ; the degrees add up
modulo $3$ under the associative multiplication. Then the algebras  $\Lambda$
and $\bar{\Lambda}$ can be merged into a bigger one if we postulate the extra
binary commutation relations between variables $\theta^A$ and
${\bar{\theta}}^{\bar{B}} \, :$
$$ \theta^A \, {\bar{\theta}}^{\bar{B}} = \omega \, {\bar{\theta}}^{\bar{B}}
\, \theta^A \, , \ \ \, \ \ \, \ \  {\bar{\theta}}^{\bar{B}} \, \theta^A =
\omega^2 \, \theta^A \, {\bar{\theta}}^{\bar{B}} \, . $$
\indent
If $A,B,C,....=1,2,...N$, then the total dimension of this algebra is
$$D(N) = 1 + 2 N + 3 N^2 + \frac{2 (N^3-N)}{3} =
\frac{2 N^3 + 9 N^2 + 4 N +3}{3} $$
These algebras are the most natural $Z_3$-graded generalizations of usual
$Z_2$-graded algebras of fermionic (anticommuting) variables.
Similarly, cubic Clifford algebras can be defined if their generators $Q^b$
are supposed to satisfy the following ternary commutation relations:
\begin{equation}
 Q^a Q^b Q^c = \omega \, Q^b Q^c Q^a + \omega^2 \, Q^c Q^a Q^b +
3 \, \rho^{abc} \, {\bf 1} \, .
\label{Clifftern}
\end{equation}
\newpage
\indent
instead of usual binary constitutive relations
$$\gamma^{\mu} \gamma^{\lambda} = (-1) \, \gamma^{\lambda} \gamma^{\mu} +
g^{\mu \lambda} \, {\bf 1} \, . $$
\indent
A {\it conjugate} ternary Clifford algebra isomorphic with the above is
readily defined if we introduce the conjugate matrices ${\bar{Q}}^{a}$
satisfying similar ternary condition with $\omega$ replacing $\omega^2$ 
and vice versa (\cite{Fleury}).
\indent
Applying cyclic permutation operator $\pi$ to all triplets of indeces on
both sides of the definition (\ref{Clifftern}), one easily arrives at the
condition that must be satisfied by the tensor $\eta^{abc}$, corresponding to
the symmetry condition on the metric tensor $g^{\mu \lambda}$ in the usual
(binary) case:
$$ \rho^{abc} + \omega \, \rho^{bca} + \omega^2 \, \rho^{cab} = 0 \, . $$
This equation has two independent solutions,
$$ \rho^{abc} = \rho^{bca} = \rho^{cab} \, , \, \ \ \, \ \ {\rm and} \, \ \
\, \ \ \, \rho^{abc} = \omega^2 \, \rho^{bca} = \omega \, \rho^{cab} \, . $$
The second, non-trivial solution defines a {\it cubic matrix} $\rho^{abc}$;
its conjugate, satisfying complex conjugate ternary relations, provides a
$Z_3$-conjugate matrix ${\bar{\rho}}^{abc}$.
\newline
\indent
These two non-trivial solutions, denoted by $\rho^{(1)}$ and $\rho^{(2)}$;
form an interesting non-associative ternary algebra with ternary
multiplication rule defined as follows (\cite{Kerner3} , \cite{Vainerman}):
\begin{equation}
( \rho^{(i)} \, * \, \rho^{(k)} \, * \, \rho^{(m)} )_{abc} =
{\displaystyle{\sum_{d, e, f}}} \, \rho^{(i)}_{fad} \, \rho^{(k)}_{dbe} \,
\rho^{(m)}_{ecf} \, .
\end{equation}
\indent
A $Z_3$-graded analogue of usual commutator as readily defined as
\begin{equation}
\{ \rho^{(i)} , \rho^{(j)} , \rho^{(k)} \} := \rho^{(i)} * \rho^{(j)} * \rho^{(k)}
+ \omega \, \rho^{(j)} * \rho^{(k)} * \rho^{(i)} + \omega^2 \,
\rho^{(k)} * \rho^{(i)} * \rho^{(j)} \; ;
\end{equation}
It has been shown in (\cite{Vainerman}) that this ternary algebra spanned by
two generators $\rho^{(1)}$ and $\rho^{(2)}$ can be represented by ordinary
matrices ( which are nothing else but two arbitratily chosen Pauli matrices)
with ternary multiplication defined as
$$ \{ \sigma^1 , \sigma^2 , \sigma^1 \} := \sigma^1 \sigma^2 \sigma^1 +
\omega \, \sigma^2 \sigma^1 \sigma^1 + \omega^2 \, \sigma^1 \sigma^1 \sigma^2
= - 2 \ \sigma^2 \, , \ \ {\rm etc.}$$
which is an illustration of the observation made by A. Sitarz (\cite{Sitarz}).
\vskip 0.4cm
\indent
{\tbf 4. CUBIC  ROOTS  OF LINEAR  DIFFERENTIAL OPERATORS.}
\vskip 0.3cm
\indent
The existence and particular properties of cubic Grassmann and Clifford
algebras suggest that they can be used in order to define {\it cubic roots}
of linear differential operators, in the same sense as the Dirac operator is
said to represent a square root of the Klein-Gordon operator, and the
generators of the supersymmetric translations are said to represent square
root of the Dirac operator.
\newline
\indent
The search for the ``cubic root'' of linear differential operator (which need
not to be the Dirac operator $\gamma^{\mu} \, \partial_{\mu} + m$) consists 
in defining  (pseudo)-differential operators whose third power, or a special
cubic combionation, will yield the linear differential operator we started
with.
\newline
\indent
Such tentatives have been made some time ago (\cite{Kerner3}, (\cite{Chung},
\cite{VARKBLR}),
and were only partially successful. Formal solutions have been found, but
their relation with the Lorentz transformations remains unclear.
\newline
\indent
The construction is based on the analogy with the supersymmetry generators,
which are realized as pseudo-differential operators acting on the fields
which are functions of the space-time variables $x^{\mu}$ and of the
anti-commuting spinorial variables $\theta^{\alpha}$ and
${\bar{\theta}}^{\dot{\beta}}$. These variables anticommute with each other
following the rule
\begin{equation}
\theta^{\alpha} \, \theta^{\beta} + \theta^{\beta} \, \theta^{\alpha} = 0,
\, \ \  \, \ \ \bar{\theta}^{\dot{\alpha}} \, \bar{\theta}^{\dot{\beta}} +
\bar{\theta}^{\dot{\beta}} \, \bar{\theta}^{\dot{\alpha}} = 0,
\, \ \ \, \ \ \, 
\theta^{\alpha} \, \bar{\theta}^{\dot{\beta}} + \bar{\theta}^{\dot{\beta}} \,
\theta^{\alpha} = 0,
\end{equation}
A formal partial derivation can be defined, satisfying the following
{\it anti-Leibniz} rule:
\begin{equation}
\partial_{\alpha} \, \theta^{\beta} = \delta^{\beta}_{\alpha} \, , \ \  
\partial_{\dot{\beta}} \, {\bar{\theta}}^{\dot{\alpha}} = 
\delta^{\dot{\alpha}}_{\dot{\beta}} \, ; \ \  \partial_{\alpha} \, 
{\bar{\theta}}^{\dot{\alpha}} = 0 \, , \, \partial_{\dot{\beta}} 
\theta^{\alpha} = 0 \, ; \, \ \ \,  \partial_{\alpha} \, ( \theta^{\beta}
\, \theta^{\gamma} ) = \delta^{\beta}_{\alpha} \, \theta^{\gamma} - 
\delta^{\gamma}_{\alpha} \, \theta^{\beta} \, , \,   {\rm etc.} \, .
\end{equation}
The generators of the supersymmetry translations are defined then as
\begin{equation}
{\cal{D}}_{\alpha} = \partial_{\alpha}  + {\bar{\theta}}^{\dot{\beta}} \,
\sigma^{\mu}_{\alpha \dot{\beta}} \,\partial_{\mu} \, , \ \ \, \ \ \, \ \
{\bar{\cal{D}}}_{\dot{\beta}} = \partial_{\dot{\beta}}  + {\theta}^{\alpha} \,
\sigma^{\lambda}_{\alpha \dot{\beta}} \,\partial_{\lambda}
\end{equation}
and are supposed to act on the space of generalized functions of space-time points
and Grassmann variables, i.e. formal {\it hermitian} series
$$ \Phi \, (x^{\mu}, \theta^{\alpha}, {\bar{\theta}}^{\dot{\beta}} ) = \phi \,
( x^{\mu}) + \psi_{\alpha} \, (x^{\mu}) \, \theta^{\alpha} + 
{\bar{\theta}}^{\dot{\beta}} \, {\bar{\psi}}_{\dot{\beta}} \, (x^{\mu}) +
W_{\mu} \, (x^{\lambda}) \, \theta^{\alpha} \,  \sigma^{\mu}_{\alpha \dot{\beta}}
\, {\bar{\theta}}^{\dot{\beta}} + ... $$
Therefore, a special quadratic combination of the supersymmetry translations
yields a linear combination of space-time translations, combined with Pauli
matrices as coefficients, which enables us to generate the full Poincar\'e
group.
\newline
\indent
The existence of ternary generalization of Grassmann variables displayed 
in the previous Section suggests how to construct the operators acting on
formal polynomial series spanned by this $Z_3$-graded algebra. Introducing
partial derivations with respect to these variables as follows:
$$\partial_A \, \theta^B = \delta^B_A \, , \, \ \ \partial_{\bar{C}} \,
{\bar{\theta}}^{\bar{D}} = \delta^{\bar{D}}_{\bar{C}} \, , \, \ \
\partial_A \, ( \, \theta^B \, \theta^C \, ) = \delta^B_A \, \theta^C +
\omega \, \delta^C_A \, \theta^B , \, \ \ \, {\rm etc.} , $$
\indent
Thus defined $Z_3$-graded derivations satisfy ternary commuutation relations
$$\partial_A \partial_B \partial_C = \omega \, \partial_B \partial_C
\partial_A \, , \, \ \ \partial_{\bar{A}} \partial_{\bar{B}} \partial_{\bar{C}}
= \omega^2 \, \partial_{\bar{B}} \partial_{\bar{C}} \partial_{\bar{A}} \, ,
\, \ \  $$
\indent
It is easy to prove (\cite{Kerner3}) that any polynomial in variables $\theta$
of order four must vanish; if one extends the commutation rules to the entire
algebra (treating, for example, all products of two $\theta$ variables,
$\theta^A \, \theta^B$, as variables of degree $2$, all the products of
the type $\theta^A \, {\bar{\theta}}^{\bar{B}}$ as variables of $Z_3$-degree
$0$, and so on, then the only surviving combinations are
$$ {\cal{A}}_1 \, : \, \ \ \{ \theta \, ,  \ \ {\bar{\theta}} \,
{\bar{\theta} \} \, , \, \ \ {\cal{A}}_2 \, : \, \ \ \{ {\bar{\theta}} \, , 
\ \ \theta \, \theta} \} \, ; \ \ \,   \, {\cal{A}}_0 \, : \, \ \
\{ \, {\bf 1} \, , \, \ \ \theta \, {\bar{\theta}} \, , \, \ \ \,
{\bar{\theta}} \, \theta \, , \, \ \ \, \theta \theta \theta \, , \,
\ \ \, {\bar{\theta}} {\bar{\theta}}  {\bar{\theta}} \, \} \, .   $$
The (pseudo)differential operators whose third powers yield a linear
differential operator should have the following form (\cite{VARKBLR}):
\begin{equation}
{\cal{D}}_A = \partial_A + \rho^{\alpha}_{A {\bar{B}} {\hat{C}}} \,
{\bar{\theta}}^{\bar{B}} \, {\hat{\theta}}^{\hat{C}} \, {\cal{D}}_{\alpha}
+ \pi^{\mu}_{A {\bar{B}}} \, {\bar{\theta}}^{\bar{B}} \, \partial_{\mu}\, ,
\end{equation}
$$ {\bar{\cal{D}}}_{\bar{B}} = \partial_{\bar{B}} +
\rho^{{\dot{\alpha}}}_{A {\bar{B}} {\hat{C}}} \,{\theta}^{A} \,
{\hat{\theta}}^{\hat{C}} \, {\bar{\cal{D}}}_{\dot{\alpha}} +
\pi^{\lambda}_{A {\bar{B}}} \, \theta^A \, \partial{\lambda} \, , $$
\begin{equation}
 {\hat{\cal{D}}}_{\hat{C}} = \partial_{\hat{C}} +
\rho^{{\dot{\alpha}}}_{A {\bar{B}} {\hat{C}}} \,{\theta}^{A} \,
{\bar{\theta}}^{\bar{B}} \, {\bar{\cal{D}}}_{\dot{\alpha}} +
\pi^{\lambda}_{A {\hat{C}}} \, \theta^A \, \partial{\lambda} \, ,
\end{equation}
where we have introduced three different types of $Z_3$-graded variables
$\theta$ , and their conjugates. The above three combinations do not exhaust
all the possibilities of construction of such operators; there are nine
other combinations.  It has been shown in (\cite{Kerner3}), (\cite{VARKBLR}),
how, under some conditions imposed on scalars (i.e. zeroth $Z_3$-degree
polynomials in $\theta$'s), special ternary or binary expressions in the
operators ${\cal{D}}_A$, ${\bar{\cal{D}}}_{\bar{B}}$ and
${\hat{\cal{D}}}_{\hat{C}}$ lead to linear expressions in supersymmetry
generators ${\cal{D}}_{\alpha}$ and ${\bar{\cal{D}}}_{\dot{\beta}}$ or in
the Poincar\'e translations $\partial_{\mu}$.
\newline
\indent
A ternary linearization of the Klein-Gordon equation has been proposed recently 
by M. Plyushchay and M. Rausch de Traubenberg (\cite{Plyushchay}. In order
to obtain a first-order differential relativistic equation of the form
\begin{equation}
\biggl( \, i \, \Gamma^{\mu} \partial_{\mu} + m \tilde{\Gamma} \, \biggr) \,
\psi := {\cal{D}} \, \psi = 0 \, ,
\end{equation}
such that ${\cal{D}}$ would satisfy
\begin{equation}
{\cal{D}}^3 \, \psi = - m \biggl( \eta^{\mu \nu} \partial_{\mu} \partial_{\nu}
 + m^2 \biggr) \, \psi \, ,
\end{equation}
where $\eta^{\mu \nu}$ denotes the Minkowskian space-time metric, $\mu,\nu =
0,1,2,3$, one has to introduce a new cubic algebra, called {\it Clifford
algebra of polynomials}, which is defined as follows. Let us denote by
$S_3(a,b,c)$ the sum of all permutations of the product $abc$,
$$S_3(a,b,c) = \frac{1}{6} \, ( abc+bca+cab+acb+cba+bac) ; $$
then we require the following identities to hold:
$$ S_3( \tilde{\Gamma}, \tilde{\Gamma}, \tilde{\Gamma}) = {\tilde{\Gamma}}^3
= - {\bf 1} \, ; \, \ \  \, S_3 (\Gamma^{\mu} , \tilde{\Gamma}, \tilde{\Gamma})
= 0 \, ;    $$
\begin{equation}
S_3 ( \Gamma^{\mu} , \Gamma^{\nu} , \tilde{\Gamma} ) = \frac{1}{3} \,
\eta^{\mu  \nu} \, , \ \ \, \ \ \, S_3 ( \Gamma^{\mu} , \Gamma^{\rho} ,
\Gamma^{\lambda} ) = 0 \, .
\end{equation}
The group of outer automorphisms of this algebraic structure is
$ SO(3,1) \times {\bf Z}_2 \times {\bf Z}_2 \times {\bf Z}_3 . $
The two factors ${\bf Z}_2 $ correspond to the $PT$-invariance of the
constitutive relations, whereas the factor ${\bf Z}_3$ is associated with
the obvious automorphism generated by the substitution
$$ (\Gamma^{\mu} \, , \, \tilde{\Gamma}) \rightarrow (\omega \, \Gamma^{\mu}
\, , \, \omega \, \tilde{\Gamma} ) $$
\noindent
With the generators $\Gamma^{\mu}$ and $\tilde{\Gamma}$, the following
Minkowskian $4$-vectors can be constructed:
$$ W^{\mu} = {\tilde{\Gamma}}^2 \, \Gamma^{\mu} \, , \, \ \ \,
{\tilde{W}}^{\mu} = \Gamma^{\mu} \, {\tilde{\Gamma}}^2 \, , \ \ \,
{\hat{W}}^{\mu} = \tilde{\Gamma} \, \Gamma^{\mu} \,\tilde{\Gamma} \, . $$
It is easy to check that one has $ W^{\mu} + {\tilde{W}}^{\mu} +
{\hat{W}}^{\mu} = 0 . $
\newline
\indent
The generatores of the Lorentz transformations have been constructed in a
few particular representations only, and with supplementary constraints. For
example:
$$J^{(1)} ={\hat{W}}_{\mu} W_{\lambda} + {\tilde{W}}_{\mu} {\hat{W}}_{\lambda}
+ {\tilde{W}}_{\mu} W_{\lambda} = \tilde{\Gamma} \Gamma_{\mu} \Gamma_{\lambda} +
\Gamma_{\mu} {\tilde{\Gamma}} \Gamma_{\lambda} + \Gamma_{\mu} \Gamma_{\lambda}
{\tilde{\Gamma}} \, ; $$
\begin{equation}
J^{(2)}_{\mu \lambda}= W_{\mu} W_{\lambda} + {\tilde{W}}_{\mu}
{\tilde{W}}_{\lambda} + {\tilde{W}}_{\mu} W_{\lambda} = {\tilde{\Gamma}}^2
\Gamma_{\mu} {\tilde{\Gamma}}^2 \Gamma_{\lambda} + \Gamma_{\mu}
{\tilde{\Gamma}}^2 \Gamma_{\lambda} {\tilde{\Gamma}}^2 + {\tilde{\Gamma}}^2
\Gamma_{\mu} \Gamma_{\lambda} {\tilde{\Gamma}}^2 .
\end{equation}
It can be shown that the operators $\frac{i}{2} \, J^{(1)}_{\mu \lambda}$
satisfy the commutation rules of the Lorentz group if the
following extra condition is imposed on the matrices $\Gamma_{\mu}$ and
$\tilde{\Gamma}$ :
\begin{equation}
\biggl[ \, ( \Gamma_{\mu} \Gamma_{\rho} \Gamma_{\lambda} + \Gamma_{\mu}
\Gamma_{\lambda} \Gamma_{\rho} + \Gamma_{\lambda} \Gamma_{\mu} \Gamma_{\rho}
\, ) \,  ,  \, \tilde{\Gamma} \biggr] = 0
\end{equation}
\indent
The obvious aim of all these constructions is to produce a model of algebraic
quark confinement.
\newpage
\indent
{\tbf 5. $Z_3$-GRADED EXTERIOR DIFFERENTIALS.}
\vskip 0.3cm
\indent
The $Z_3$-graded ternary analogue of Grassmann algebra suggests the existence
of a generalized exterior differential calculus based on cubic commutation
relations. As a matter of fact, such differential calculus has been developed
in a series of papers in the 90-ties (\cite{Kerner3} , \cite{Kerner5} ).
\newline
\indent
The starting point for the introduction of exterior differentials can be
the observation that while the first differentials of local coordinates on
a manifold transform naturally as vectors, this does not remain true for
the second and higher order differentials. As a matter of fact, consider
formal first, then second-order differentials of a function $f (x^k)$
defined on a manifold with local coordinates $x^k$ :
\begin{equation}
df = (\partial_k \, f) \, dx^k \, ; \, \ \ d^2 f = (\partial^2_{i k } \, f )
\, d x^i \, dx^k + (\partial_k \, f ) \, d^2 x^k \, , \, \ \ 
\label{onetwo}
\end{equation}
It becomes obvious that in order to ensure the nilpotency of the operator $d$,
i.e. $d^2 = 0$, one has to assume that the product of $1$-forms $d x^i$ is
antisymmetric, $d x^i \wedge d x^k = - d x^k \wedge d x^i$. However, if this
condition is not imposed, then $d^2 x^k \neq 0$, and it combines with the
symmetric part of the product $d x^i \, dx^k$. Then it is not difficult to
impose third-order nilpotency, $d^3 = 0$.
\newline
\indent
Let $M$ be a smooth $n$-dimensional manifold and let $\omega$ be a $3$-rd
primitive root of unity, $\omega = e^{2 \pi i/3}, \omega^3 = 1 . $  Let $U$
be an open subset of $M$ with local coordinates  $x_1, x_2, \ldots, x_n$.
Our aim is to construct an analogue of the exterior algebra of differential
forms with exterior differential $d$ satisfying the $\omega$-Leibniz rule 
\begin{equation}\label{Leibniz}
d(\phi \theta)=d\phi \,\theta + \omega^{\vert \phi \vert}\,\phi \,d \theta,
\end{equation}
where $\phi , \theta$ are complex valued differential forms, 
$\vert \phi \vert$ is the degree of $\phi$, and $d^3=0 .$  We shall also
assume that as in the usual $Z_2$-graded case, $d$ is a linear operator
that raises the degree of any exterior form by one.
\newline
\indent
As in the usual exterior differential calculus, we identify the vector space 
of differential forms of degree zero, denoted by $\Omega^0(M)$, with the 
space of smooth functions on the manifold $M$. We shall assume that there is 
no difference between the vector space of differential forms of degree one in 
our case (denoted by $\Omega^1(M)$) and the same vector space in the case of 
the classical exterior algebra. Thus
$  \Omega^1(M) = \{ \, \omega_i\,dx^i \} :  \, \ \ {\rm where} \, \ \
\omega_i = \omega_i (x^k) , \, \, i=1, 2, \ldots, n\;
\mbox{are smooth functions on $M$ } \, .$
\newline
\indent
The assumption $d^2 \not =0$ implies that there is no reason
to use only the first order differentials $dx_i$ in the construction of the 
algebra of differential forms induced by $d$; one can also add a set of 
formal {\it second-order differentials}, in which case the algebra will be 
generated by  
$$dx^1, \dots, dx^n, \dots, d^2 x^1, \dots, d^2 x^n .$$
In order to endow the algebra of differential forms with appropriate
$Z_3$-grading we shall associate the degree $k$ to each differential 
$d^k x^i$.  As usual, the degree of the product of differentials is the sum 
of the degrees of its components modulo $3$. Given any smooth function $f$ 
and successively applying to it the exterior differential $d$ one obtains 
the following expressions for the first three steps: 
\begin{equation}\label{firstdifferential}
d f= (\partial_i f )\; dx^i,  \, \ \ \, \ \
d^2 f= (\partial^2_{ij} f ) \; dx^{(i} dx^{j)} + (\partial_i  f) \; d^2 x^i ,
\end{equation}
\begin{equation}
\label{thirddifferential}
d^3 f=(\partial^3_{ijk} f ) \; dx^{(i} dx^j dx^{k)} + (\partial^2_{ij} f )
\; (d^2 x^i, dx^j)_{\omega} + (\partial_i f ) \; d^3 x^i.
\end{equation}
Because the partial derivatives of a smooth function do commute, only the
totally symmetric combinations of indices are relevant here. This is why
in the above formula the parentheses mean the symmetrization
with respect to the superscripts they contain, i.e. 
\begin{equation}
dx^{(i} dx^{j)}={1\over 2!}\, (dx^i\,dx^j + dx^j\,dx^i),
\end{equation}
\begin{equation}
dx^{(i} dx^j dx^{k)} = \frac{1}{3!} \,
{\displaystyle{\sum_{\pi \in {\cal{S}}_3}}}
dx^{\pi(i)} dx^{\pi(j)} dx^{\pi(k)}
\end{equation}
\begin{equation}
 \, \ \ {\rm and \, } \, \ \ (d^2 x^i, dx^j)_{\omega}=d^2 x^{(i}\, dx^{j)} + 
(1+ \omega)\;d x^{(i} d^2 x^{j)}.
\end{equation}
\indent
In order to guarantee that $d^3 f = 0$ for any smooth function on $M$, the
following three conditions have to be satisfied:
\begin{equation}
\label{N=3}
dx^{(i} dx^j dx^{k)}=0,\, \ \ \, 
d^2 x^{(i} dx^{j)}+(1+ \omega) d x^{(i} d^2 x^{j)}=0,\, \ \ \, \ \ 
d^3 x^i = 0.
\end{equation}
These relations represent the minimal set of conditions that should be
imposed on the differentials in order to ensure $d^3 = 0 .$ From the first
condition it is obvious that first differentials are always $3$-nilpotent,
$(dx^k)^3 = 0 .$ On the other hand the equations (\ref{N=3}) demonstrate
clearly that generally there are no relations
implying the nilpotency of any power for the differentials of higher order. 
Therefore though the algebra generated by the relations (\ref{N=3}) is 
finite-dimensional with respect to the first order differentials because of 
$(dx^k)^3 = 0$, it remains infinite-dimensional with respect to the 
entire set of differentials. 
\newline
\indent
We solve the first condition in
(\ref{N=3}) by assuming that each cyclic permutation of any three 
differentials of first order is accompanied by the factor $\omega$ which in
this case is a primitive cubic root of unity and satisfies the identity
$1+\omega+\omega^2=0.$
\newline
\indent
Thus we assume that each triple of differentials of first order
$dx^i, dx^j, dx^k$ is subjected to {\it ternary} commutation relations
\begin{equation}\label{ternary}
dx^i dx^j dx^k = \omega \; dx^j dx^k dx^i.
\end{equation}
These ternary commutation relations can not be made compatible with binary
commutation relations of any kind. 
Therefore we suppose that all binary products $dx^i dx^j$ are independent 
quantities. The second condition in (\ref{N=3}) can be easily solved by 
assuming the following commutation relations:
\begin{equation}
\label{binary}
dx^i d^2 x^l= \omega \; d^2 x^l dx^i.
\end{equation}
Note that from (\ref{ternary}) and (\ref{binary}) it follows that the above
ternary and binary commutation relations are coherent with the $Z_3$-grading,
i.e. the quantities $d x^k d x^m $ and
$d^2 x^j$ behave as elements of degree $2$ and could be interchanged in the
formulae (\ref{ternary}) and (\ref{binary}) .
\newline
\indent
The ternary commutation relations (\ref{ternary}) are much stronger than the 
cubic nilpotence which follows from the first relation of (\ref{N=3}). It has 
been proved in (\cite{Kerner3}) that if the generators of an associative 
algebra obey ternary commutation relations such as (\ref{ternary}) then all 
the expressions containing four generators should vanish. This means that the 
highest degree monomials which can be made up of the first order differentials 
have the form $dx^i dx^j dx^k\,, dx^i (dx^j)^2$. 
In order to construct an algebra with self-consistent structure we shall 
extend this fact to the higher order differentials supposing that {\it all} 
differential forms of fourth or higher degree vanish. 
\newpage
\indent
If we assume that the functions commute with the first differentials, i.e. if 
$$ x^k d x^m = d x^m x^k ,$$ 
then by virtue of the $\omega$-Leibniz rule 
the second order differentials do not commute with smooth functions, because
we get by differentiating the above equality we obtain
$$ d (x^k d x^m) = d x^k d x^m + x^k d^2 x^m = d ( d x^m x^k ) = 
d^2 x^m x^k + \omega d x^m d x^k $$
which leads to the identity
\begin{equation}
x^k d^2 x^m - d^2 x^m x^k = \omega \, ( d x^k d x^m - \omega^2 \, dx^m d x^k)
\end{equation}
In what follows, we shall consider only the expressions in which the forms
of different degrees are multiplied on the left by smooth functions of the
coordinates $x^k$, which means that we consider the algebra $\Omega (M)$ 
as a free finite-dimensional left module over the algebra of smooth functions. 
\newline
\indent
It is quite easy to evaluate the dimension of the module $\Omega (M)$ ,
which is ${\cal N} = (n^3 + 6 n^2 + 5 n)/3.$
\newline
\indent
This $Z_3$-graded of exterior differential calculus has been also realized
in other representations, among others, as a differential algebra of operators
acting on a generalized Clifford algebra (\cite{VARK}), or in other matrix
representations; a covariant formulation of this calculus, including
naturally the notions of generalized connections and curvatures, has been
elaborated recently (\cite{Kerner5}, \cite{Kerner6}, \cite{VARK}
\cite{Coquereaux}.
\newline
\indent
The homological content of the theory becomes richer than in ordinary case,
because now one can define not only the spaces $Ker(d)$, $Im(d)$, but also
$Ker(d^2)$ and $Im(d^2)$, with obvious inclusions $Im(d^2) \subset Im(d)$
and $Ker(d) \subset Ker(d^2)$, and various quotients of those; for the general
case of differential calculus based on the postulate $d^N = 0$, the full
theory is exposed in \cite{MDVRK1} \cite{MDV1}, \cite{Samani})
\newline
\indent
An interesting application of these cohomologies has been recently found
by M. Dubois-Violette and I.T. Todorov (\cite{MDVIT}) in relation with the
WZNW model and a generalization of the corresponding
BRS-symmetry operator $A$ satisfying $A^h = 0$ with $h = 2n+1, n=1,2,...$
\newline
\indent
An alternative way of realizing exterior differential calculus with $d^3 = 0$
has been proposed by M. Dubois-Violette and M. Henneaux (\cite{MDV2}).
Instead of $Z_3$-grading, one
considers all possible tensor fields whose Young diagrams have no more than
{\it two} columns. By differentiating these fields and then using the
appropriate symmetrization procedure, we can define a coherent differential
calculus with $d^3 = 0$, which may prove useful in handling higher spins,
in particular, the graviton field. As a matter of fact, in order to arrive
at physically relevant field, represented in General Relativity by Riemann
tensor, starting from the metric field $g_{\mu \nu},$ we have to differentiate
{\it twice}. Subsequently, the field equations can be cast into the form of
$d^3 (g) = 0$.
\newline
\indent
We have given here a very shortened overview of ``ternary ideas'' in
Mathematical Physics; we believe that many interesting applications are
still ahead of us.

\vskip .4cm
\indent
{\tbf Acknowledgments}
\vskip .2cm
It is a pleasure to express my thanks V. Abramov, M. Dubois-Violette,
P.P. Kulish and A. Sitarz for many valuable discussions and remarks;
particularly to P.P. Kulish for his careful reading of the manuscript.


\newpage

\begin{thebibliography}{50}

\bibitem{Cayley} A. Cayley , Cambridge Math. Journ. {\bf 4}, p. 1 (1845)

\bibitem{Sylvester} J.J. Sylvester, Johns Hopkins Circ. Journ., {\bf 3},
p.7 (1883).

\bibitem{Kapranov} M. Kapranov, I.M. Gelfand, A. Zelevinskii, {\it Discriminants,
Resultants, and Multidimensional Determinants}, Birkh\"auser ed., (1994)

\bibitem{Vainerman} L. Vainerman, R. Kerner, Journal of Math. Physics,
{\bf 37} (5), p. 2553 (1996)

\bibitem{Himbert} A. Himbert, Comptes Rendus de l'Acad.Sci. Paris, (1935);
see also : R. Kerner, {\it ``The Cubic Chessboard''}, Class. and Quantum Gravity,
{\bf 14} 1A, p. A203 (1997)

\bibitem{Dabrowski} L. Dabrowski, F. Nesti, P. Siniscalco, Int. Journ. Mod.
Phys. A 13, p. 4147 (1998)

\bibitem{Plyushchay} M. Plyushchay, M. Rausch de Traubenberg, Phys. Lett. B,
{\bf 477}, p. 276  (2000)

\bibitem{Nambu1} J. Nambu, Physical Review D {\bf 7}, p.2405 (1973)

\bibitem{Takhtajan} L. Takhtajan, Comm. Math. Physics, {\bf 160}, p. 295 (1994)

\bibitem{Yamaleev1} R.M. Yamaleev, JINR preprints JINR-E2-88-147  (1988),
JINR-E2-89-326 (1989)

\bibitem{Okubo1} S. Okubo, Journ. of Math. Physics, {\bf 34}, p. 3273;
{\it ibid ,} p. 3292  (1993)

\bibitem{Lipkin} H.J. Lipkin, {\it Frontiers of the Quark Model}, Weizmann
Inst. pr. WIS-87-47-PH (1987)

\bibitem{Sitarz} A. Sitarz, private communication (1998); also as a preprint
math.RA/9807019 .

\bibitem{Fleury} N. Fleury, M. Rausch de Traubenberg, R. Yamaleev, Int. J.
Mod. Physics A 10, p. 1269 (1995)

\bibitem{Kerner3} R. Kerner, Comptes Rendus Acad. Sci. Paris, {\bf 312}, ser. II,
p. 191 (1991)

\bibitem{Chung} W.-S. Chung, Journ. of Math. Phys. {\bf 35} (5), p. 2497 (1994)

\bibitem{VARKBLR} V. Abramov, R. Kerner, B. Le Roy, Journal of .Math.Phys.
{\bf 38} (3), 1650-1669 (1997).

\bibitem{Kerner5} R. Kerner, Journal of Math. Phys. {\bf 33} (1), p. 403 (1992)

\bibitem{VARK} V. Abramov, R. Kerner, Journal of Math. Physics, {\bf 41} (8),
p. 5598 (2000).

\bibitem{Kerner6} R. Kerner, B. Niemeyer, Lett. in Math. Phys., {\bf 45},
p.161 (1998)

\bibitem{Coquereaux} R. Coquereaux, Lett. in Math. Physics, {\bf 42}, p.241
 (1997)

\bibitem{MDVRK1} M. Dubois-Violette, R. Kerner, Acta Math. Univ. Comenianae,
{\bf LXV}, p. 175 (1996)

\bibitem{MDV1} M. Dubois-Violette, Contemp. Math. {\bf 214}, p. 69 (1998)

\bibitem{Samani} K. Samani, A. Mostafazadeh, hep-th 0007009 (2000) 

\bibitem{MDVIT} M. Dubois-Violette, I.I. Todorov, Letters in Math. Physics
{\bf 42} (2), p. 183 (1997)

\bibitem{MDV2} M. Dubois-Violette, M. Henneaux, Lett. in Math. Phys. {\bf 49}
(3), p. 245 (1999)


\end{thebibliography}
\end{document}